\documentclass{amsart}
\usepackage{amsmath}
\usepackage{amssymb}
\usepackage{mathrsfs}
\usepackage{mathtools}
\newcommand{\beq}[1]{\begin{equation}\label{#1}}
\newcommand{\eeq}{\end{equation}}
\begin{document}
%	------------------------------------------------------------
%	\\\\\\\\\\\\\\\\\\\\\\\\\\\\\\\\\\\\\\\\\\\\\\\\\\\\\\\\\\\\
%	------------------------------------------------------------
\author[J.~Fuentes]{J.~Fuentes}
\address[J.~Fuentes]{Department of Physics, CINVESTAV--IPN\\ AP 14--740, Mexico City.}
\email[J.~Fuentes]{jfuentes@fis.cinvestav.mx}
%	------------------------------------------------------------/
\author[P.~Galaviz]{P.~Galaviz}
\address[P.~Galaviz]{Department of Physics, CINVESTAV--IPN\\ AP 14--740, Mexico City.}
\email[P.~Galaviz]{pablo.galaviz@icloud.com}
%	------------------------------------------------------------/
\author[T.~Matos]{T.~Matos}
\address[T.~Matos]{Department of Physics, CINVESTAV--IPN\\ AP 14--740, Mexico City.}
\email[T.~Matos]{tmatos@fis.cinvestav.mx}
%	------------------------------------------------------------
%	\\\\\\\\\\\\\\\\\\\\\\\\\\\\\\\\\\\\\\\\\\\\\\\\\\\\\\\\\\\\
%	------------------------------------------------------------
\title{Pseudo Spectral Transform for Schr\"odinger--Poisson Equations}
\maketitle

%	  __   ____  ____  ____  ____   __    ___  ____ 
% 	 / _\ (  _ \/ ___)(_  _)(  _ \ / _\  / __)(_  _)
%	/    \ ) _ (\___ \  )(   )   //    \( (__   )(  
%	\_/\_/(____/(____/ (__) (__\_)\_/\_/ \___) (__)

\begin{abstract}

Here we present exact, stationary, parametric solutions to the Schr\"odinger--Poisson system. We confront two images: on one hand, we draw on the homotopy analysis method which leads us to a nonlinear integral scheme. Indeed, this approach might be simplified by looking for sufficiently smooth solutions vanishing asymptotically. However, since our system possesses stiffness an additional analysis has to be considered. On the other hand, we seek for exact solutions over  the inverse  scattering  method by introducing a pseudo spectral transform. In fact, this pseudo spectral method generalises Korteweg--de Vries family's kernel and let us to circumvent some technical difficulties originally arisen in our first approach although, again, we come to an integral representation, which we test for convergence.

\end{abstract}

%	 __  __ _  ____  ____   __  
%	(  )(  ( \(_  _)(  _ \ /  \ 
% 	 )( /    /  )(   )   /(  O )
%	(__)\_)__) (__) (__\_) \__/

\section{Introduction}
\label{s:intro}

%Astrophysics
% |--------------------------------------------------------------------|
The Schr\"odinger--Poisson system is  relevant  for  several  fields in  physics  and
mathematics. There are cosmological models where the dark matter takes
the form of a scalar field.  Such field dynamics for galactic halos is
described  by a Schr\"odinger--Poisson system  \cite{MatUre01, AlcGuzMat02,  GuzUre04,
  BerGuz06,  BerGuz06a}.  Hypothetical  dense  scalar  field known  as
Boson  stars  are the  relativistic  counterpart  of a  galactic  halo
\cite{Sei90c,  BalSeiSue97,   BalComShi98,  AlcBecGuz03,  AlcGonSal04, AlcDegNun10,  RuiDegAlc12}.   In the astrophysics realm, Schr\"odiger
equation  describes the dynamics  of  a  scalar field  while  Poisson
equation dictates the dynamics of the gravitational field.
% |--------------------------------------------------------------------|
%Quantum mechanics
On the other hand, in quantum mechanics and semiconductor theories the
Schr\"odinger--Poisson system models  the interaction of charged  particles in crystals
\cite{ReiGud12,  LipKolMor01}, {\em e.g.}.  Schr\"odinger
equation  describes  a  particle  dynamics under  the  presence  of  a
electric field given by Poisson equation.

%Type of solutions
% |--------------------------------------------------------------------|
A  convenient  way  to  solve  the Schr\"odinger--Poisson  equations  is  by  numerical
computations  \cite{SchBecRuc14}.   Although  there  are  several
theoretical  studies showing  some of  the general  properties of  the
system, in particular its solitonic nature  \cite{BelSic11,   IanVai12,  MelMan00}.

% |--------------------------------------------------------------------|

% |--------------------------------------------------------------------|
The first techniques to solve some celebrated solitonic equations were
laid soon after Zabusky and Kruskal reported their observations on the
Fermi--Pasta--Ulam   puzzle   \cite{zabusky65}    (even   Seeger   et
al. noticed quite the same over a decade before \cite{seeger53}).
% |--------------------------------------------------------------------|
Among  these  techniques,  the {\em  spectral  transform},  originally
introduced by Gardner  {\em et al.}  \cite{gardner67}, is  considered as the
cornerstone of soliton theory given it  is an extension of the Fourier
transform to the nonlinear framework \cite{ablowitz74}.
% |--------------------------------------------------------------------|
The theory  became augmented  by relevant enhancements  and additional
developments  to the  spectral  transform \cite{lax68,zakharov71}  as
well  as the  introduction of  alternative approaches  founded on  the
Darboux            and           B\"acklund            transformations
\cite{hirota71,estabrook76,harrison78,matveev79,levi80}.  Today,  the
length  of these  methods is  such  that these  reach applications  of
indubitable   interest  in   other  areas   of  mathematical   physics
\cite{bogdan}.
% |--------------------------------------------------------------------|

% |--------------------------------------------------------------------|
In this work we concentrate on the spectral transform in order to seek
for   exact,  stationary   solutions  to   the  Schr\"odinger--Poisson
equations. Foremost,  our procedure  is entirely heuristical  and lies
substantially on integral equations.
% |--------------------------------------------------------------------|

% |--------------------------------------------------------------------|
The paper  is organized as follows:  In section \ref{s:approximations}
approximated  solutions are  obtained  through  the homotopy  analysis
method  \cite{liao04}. This  algorithm enable  us to  solve nonlinear
integral equations, being an analogous  path to the inverse scattering
method     where      the     linear     integral      equation     of 
Gelfand--Levitan--Marchenko (GLM) raises (section
\ref{s:exact}).  Finally,  our  results   are  summarized  in  section
\ref{s:detach}.   The  presented   solutions  are   single--solitonic,
{\em i.e.}  are uniquely related to an individual bounded state.
% |--------------------------------------------------------------------|

\subsection{Units and Notation}
\label{s:notation}

In order to simplify the writing, we use dimensionless quantities along the document, unless it is stated otherwise. In particular, we use geometrical units in the very first part of section \ref{s:approximations} to introduce the Einstein--Klein--Gordon equations. From section \ref{s:exact}, Schr\"odinger equation appears in its canonical form, where $\hbar=1$. 

We try to avoid all functional dependences unless these are absolutely necessary because of ambiguity or lack of context. Also, the following notation is employed: 
\begin{itemize}
\renewcommand\labelitemi{ }
\item $\mathrm{i}$ is the imaginary unit ($\sqrt{-1}$),
\item $x,y$ are real independent variables,
\item $\iota,\kappa,\mu,\nu$ are discrete (integer) indexes (quantities),
\item $u=u(x)$ is an arbitrary test function,
\item $u_x$ denotes the derivative of the function $u$ with respect to $x$,
\item $R_{\mu\nu}$ is a tensor,
\item $K(\cdot)$ is a kernel,
\item $K_\mu(\cdot)$ is an iterated kernel,
\item $\psi, \Psi$ are wave functions,
\item $S[u]$ is the spectral transform of the function $u$,
\item $\partial_x$ denotes a partial differential operator acting on the independent variable $x$,
\item $\bigtriangleup$ stands for the Laplace operator,
\item ${\bf H}$ is the Schr\"odinger operator,
\end{itemize}
unless a different thing is specified on the text. 

%	 ____  __  ____  ____  ____ 
%	(  __)(  )(  _ \/ ___)(_  _)
%	 ) _)  )(  )   /\___ \  )(  
%	(__)  (__)(__\_)(____/ (__)

\section{First approximations}
\label{s:approximations}

% |--------------------------------------------------------------------|
The Schr\"odinger--Poisson equations
% |--------------------------------------------------------------------|
\begin{subequations}\label{sp}
  \begin{align}
    \left[- \bigtriangleup  + \, \phi\right]\psi &= \mathrm{i}\, \psi_t, \label{spa}\\
    \bigtriangleup\phi &= \vert\psi\vert^2, \label{spb}
  \end{align}
\end{subequations}
% |--------------------------------------------------------------------|
are the canonical representation of the Schr\"odinger--Newton equations originally introduced in \cite{ruffini69}, which, constitute the weak field
limit of Einstein--Klein--Gordon equations
% |--------------------------------------------------------------------|
\beq{ekg}
R_{\mu\nu} - \frac{1}{2}g_{\mu\nu}R = 8\pi\langle\psi\vert {\bf T}_{\mu\nu} \vert\psi\rangle,
\eeq
% |--------------------------------------------------------------------|
here $R_{\mu\nu}$, $g$ and ${\bf T}_{\mu\nu}$ stand for the Ricci tensor, the  metric and  the energy--momentum  operator, all in geometric units. 
% |--------------------------------------------------------------------|

% |--------------------------------------------------------------------|
From now  on we restrict our  study to one spatial  coordinate, namely
$x$. Additionally,  we will  assume stationary  solutions of  the form
$\psi(x,t) = u(x) \exp( \pm \mathrm{i} \omega t)$, where $\omega$ is a
suitable  real constant,  $t$ represents  the time  and $u:\mathscr{X}
\rightarrow \mathscr{Y}$  is, in  general, a holomorfic  function such
that   $\mathscr{X}$   and   $\mathscr{Y}$   are   open   subsets   in
$\mathbb{C}$. Thence, the nonlinear system in \eqref{sp} is reduced to
% |--------------------------------------------------------------------|
\begin{subequations}\label{spx}
  \begin{align}
    \left[-\partial^2_{x} +  b\, \phi\right]u & = \mp\, \omega\, u, \label{spxa}\\
    \phi_{xx}  & = \vert u \vert^2,\label{spxb}
  \end{align}
\end{subequations}
% |--------------------------------------------------------------------|
having  introduced a  complex  parameter $b$, for generality. The choice in the sign related to $\omega$ is physically irrelevant. Also, notice that $u$ and $\phi$ are time independent.
% |--------------------------------------------------------------------|

% |--------------------------------------------------------------------|
Broadly speaking, $u$ can be computed  in the neighbourhood $\vert x -
x_0\vert$ by  Pad\'e approximants of the form $[\mu/\mu+1]_{U}(x)$,
with   $U   =   u_x/u$   (furthermore details can be found in \cite{baker73,brezinski80}).   Yet,   first
approximations of $u$ can be obtained by un-coupling \eqref{spx}.
% |--------------------------------------------------------------------|

% |--------------------------------------------------------------------|
Consider \eqref{spxa}, which differentiated  two times with respect to
$x$ reads
% |--------------------------------------------------------------------|
\beq{schxx}
-u_{xxxx}  +  b\,(\phi\,  u_{xx}   +  \phi_{xx}\,u)  +  2\,b\,\phi_{x}
\,u_{x} = \omega\, u_{xx},
\eeq
% |--------------------------------------------------------------------|
by  making the  substitutions  $\phi_{xx}\rightarrow\vert u  \vert^2$,
$(b\phi  -  \omega)u\rightarrow  u_{xx}$  and  solving  for  $\phi_x$,
\eqref{schxx} becomes
% |--------------------------------------------------------------------|
\beq{usch}
\phi_{x} = \int_{\mathscr{X}}\mathrm{d}{x}\, \vert u \vert^2 = \frac{u_{xxxx}-u^{-1}(u_{xx})^2-b\,u\,\vert u\vert^2}{2\,b\,u_{x}}.
\eeq
% |--------------------------------------------------------------------|

% |--------------------------------------------------------------------|
Yet again,  consider \eqref{spxa}, by differentiating  it with respect
to $x$, solving for $\phi_x$ and equating with \eqref{usch}, we obtain
an     uncoupled     scheme      of     the     Schr\"odinger--Poisson
equations:
% |--------------------------------------------------------------------|
\beq{spunc}
u_{xxxx} = \frac{b\,\vert u \vert^2\,u^3 + [\,(u_{xx})^2+2\,u_{x}
    \,u_{xxx}\,]\,u-2\,(u_{xx})^2\,u_{x}}{u^2},
\eeq
% |--------------------------------------------------------------------|
that possess just the same integral curves than \eqref{spx}.
% |--------------------------------------------------------------------|

% |--------------------------------------------------------------------|
In particular, if we consider $u$ as a real function to be regular for
all real values of $x$ and to vanish asymptotically rapid, viz.
% |--------------------------------------------------------------------|
\beq{levi}
\lim_{x\rightarrow\pm\infty} \left[\, \vert x \vert^{\epsilon+1}\,u(x)\,\right] = 0,\quad 0 < \epsilon, 
\eeq
% |--------------------------------------------------------------------|
only the first term in the  right--hand side of eq. \eqref{spunc} does
contribute significantly if compared to the others, consequently
% |--------------------------------------------------------------------|
\beq{spux}
u_{xxxx} \approx b\,u^3,
\eeq
% |--------------------------------------------------------------------|
which general Urysohn form is
% |--------------------------------------------------------------------|
\beq{spu}
\begin{split}
u(x) = C_1\frac{b x^3}{6} & \int^x\mathrm{d}y\, u^3(y) - C_2\frac{b x^2}{2}  \int^x\mathrm{d}y\, y \, u^3(y) \\ 
 + C_3\frac{b x}{2} & \int^x\mathrm{d}y\, y^2\, u^3(y) - C_4\frac{b}{6} \int^x\mathrm{d}y\, y^3\, u^3(y) + F(x),
\end{split}
\eeq
% |--------------------------------------------------------------------|
here $C_i$ are constansts determined  from boundary conditions and the
function $F(x)$ equals  to $c_1 + c_2 x  + c_3 x^2 + c_4  x^3 + f(x)$,
where $c_i$ stand  for constants and $f=f(x)$ is a
real  valued function  to be  continuous  in the  sense of  Lipschitz, to assure that 
equation \eqref{spu} has a unique solution.
% |--------------------------------------------------------------------|

% |--------------------------------------------------------------------|
In  order  to  approximate  the solution  of  the  nonlinear  integral
equation  \eqref{spu},  hither  we  focus  on  the  homotopy  analysis
method.  That being  said, assume  that  each individual  term of  the
Urysohn  form \eqref{spu}  has  a  kernel $K(x,y)\in\mathbb{R}$.   Let
$\Xi_\mu(u^3)\in\mathbb{R}$   be  the   $\mu$--th  order   deformation
constraint given by
% |--------------------------------------------------------------------|
\beq{deformation}
\Xi_\mu(u^3) = \sum_{\nu=0}^{\mu}u_{\mu-\nu}\sum_{\beta=0}^{\nu}u_{\beta}u_{\nu-\beta}, 
\eeq
% |--------------------------------------------------------------------|
then, the homotopy--series solution to \eqref{spu} is computed through
% |--------------------------------------------------------------------|
\beq{homo}
u(x) = \sum_{\mu=0}^{\infty} u_\mu(x) = \sum_{\mu=0}^{\infty}\left\{ \int^x\mathrm{d}y\,K(x,y)\,\Xi_{\mu-1}(u^3)\right\},
\eeq
% |--------------------------------------------------------------------|
subject to the initial guess $u_0 = f$.
% |--------------------------------------------------------------------|
The  last  expression is  a  pivotal  trace  in  our quest  for  exact
solutions. A  way to  obtain general  exact solutions  to \eqref{spx}  is by  the
inverse scattering method, which is the spanned version of \eqref{homo}. See our discussion
in  section  \ref{s:exact}.
% |--------------------------------------------------------------------|
In  our  current  analysis,  while   the  kernel  in  \eqref{homo}  is
separable,  given the  interval $x\in[\chi_1,\chi_2]$,  convergence of
series \eqref{homo} is assured iff
% |--------------------------------------------------------------------|
\beq{conver}
\left\vert 1 - \int_{\chi_1}^{\chi_2}\mathrm{d}y\, K(y,y) \right\vert \leq 1,
\eeq 
% |--------------------------------------------------------------------|
is  undeniably   satisfied\footnote{Cf.  \cite{liao,awawdeh}   for  a
  comprehensive  discussion.}. Thus,  at  the hand  of  allowing $f  =
\mathrm{sech}(a x)$, with  $a$ some real constant, it  does imply that
uniquely  the $C_2$  and $C_4$  terms  in \eqref{spu}  will behave  as
required by \eqref{levi}, so the constants $c_1=c_2=c_3=c_4=0$ and the
solution to \eqref{spu} is approximated in virtue of
% |--------------------------------------------------------------------|
\beq{sol}
u(x) = \mathrm{sech}(a x) - b \sum_{\mu=1}^\infty \left\{ \frac{C_2x^2}{2}  \int^x\mathrm{d}y\, y\, \Xi_{\mu-1}(u^3) + \frac{C_4}{6}  \int^x\mathrm{d}y\, y^3\, \Xi_{\mu-1}(u^3)\right \};
\eeq
% |--------------------------------------------------------------------|
indeed, the programme \eqref{sol} lead  to  the  exact  solution  to  \eqref{spu}, iff the series  \eqref{homo} does  converge. Unfortunately, these integrals may become complicated, hence, in most cases, \eqref{sol}  should be enough  for numerical computation  of the
first rough  solutions.

% |--------------------------------------------------------------------|
Moreover,  the  fact  that  we  have restricted  our  study  to  those
solutions  compelled  to  suffice  \eqref{levi}, is  only  a  feasible
gimmick  to work  for  solutions of  accesible  interpretation in  the
physics  environment.   Nonetheless,  a   different  class   might  be
considered for another purposes.
% |--------------------------------------------------------------------|

% |--------------------------------------------------------------------|
In  the  next  section  we  study the  generalization  of  the  former
blueprint in accordance with a parallel essay from soliton theory.
% |--------------------------------------------------------------------|

%	 ____  ____  _  _  __  ____  ____ 
%	(    \(  __)/ )( \(  )/ ___)(  __)
%	 ) D ( ) _) \ \/ / )( \___ \ ) _) 
%	(____/(____) \__/ (__)(____/(____)

\section{Devising exact solutions}
\label{s:exact}

% |--------------------------------------------------------------------|
Straightaway, looking at the stationary Schr\"odinger equation defined
in all $x\in\mathbb{R}$
% |--------------------------------------------------------------------|
\beq{schro}
{\bf H}\Psi = k^2 \Psi,
\eeq
% |--------------------------------------------------------------------|
the discrete  part of  the operator  ${\bf H}  = -\partial_{xx}  + u$,
consists  of  $A$  negative  eigenvalues  $k^2  =  -p^2_\alpha$,  with
$p_\alpha>0$, $\alpha  = 1,2,\ldots,A$.  To each of  these eigenvalues
corresponds a  unique solution  $\Psi_\alpha$ impelled to  satisfy the
boundary condition:
% |--------------------------------------------------------------------|
\beq{bnd1}
\lim_{x\rightarrow\infty} [\,\exp(p_\alpha x)\, \Psi_\alpha(x)\,] = 1, \quad \alpha = 1,2,\ldots,A.
\eeq
% |--------------------------------------------------------------------|

% |--------------------------------------------------------------------|
Whereas the  continuum part, featured  by all positive real  values of
the eigenvalue  $k^2$, does typify  the solution $\Psi$  to accomplish
the asymptotic boundary conditions:
% |--------------------------------------------------------------------|
\beq{bnd2}
\begin{split}
\Psi &\rightarrow T(k) \exp(-\mathrm{i}kx), \quad x\rightarrow-\infty,\\
\Psi &\rightarrow \exp(-\mathrm{i}kx) + R(k) \exp(\mathrm{i}kx), \quad x\rightarrow+\infty,
\end{split}
\eeq
% |--------------------------------------------------------------------|
where  $T(k)$ and  $R(k)$ stand  for the  transmission and  reflection
coefficients\footnote{To behold how both \eqref{bnd1} and \eqref{bnd2}
  are certainly consistent with \eqref{levi}.}.
% |--------------------------------------------------------------------|

% |--------------------------------------------------------------------|
Having  said all  that, the  extended  spectral transform  $S$ of  the
function $u$ is defined as
% |--------------------------------------------------------------------|
\beq{st}
S[u] := \{R(k),\; -\infty<k<\infty;\,p_\alpha, \; q_\alpha,\; \alpha = 1,2,\ldots,A,\, \beta\in\mathbb{Z}\,\},
\eeq
% |--------------------------------------------------------------------|
where  $q_\alpha$  are the  normalization  factors  and $\beta$  is  a
discrete parameter related to the GLM equation
% |--------------------------------------------------------------------|
\beq{glm}
W(x,y) + K(x+y) + \int_x^\infty\mathrm{d}z W(x,z)K(z+y)=0,\quad y>x,
\eeq
% |--------------------------------------------------------------------|
through the complex valued function   
% |--------------------------------------------------------------------|
\beq{glmk}
K(x) = \frac{1}{2\pi} \int \mathrm{d}k \; \exp(\mathrm{i}kx)R(k) + \sum_{\alpha=1}^{\alpha} q_{\alpha} \,\Gamma\left[\frac{1}{\beta + 1}, p_\alpha (-x)^{\beta+1}\right],
\eeq
% |--------------------------------------------------------------------|
where
% |--------------------------------------------------------------------|
\begin{equation*}
\Gamma(a,z) := \int_{z}^{\infty}\mathrm{d}s \, \exp(-s)s^{a - 1},
\end{equation*}
% |--------------------------------------------------------------------|
is the so called {\em incomplete gamma function}.
% |--------------------------------------------------------------------|

% |--------------------------------------------------------------------|
We  request the  extended  spectral transform  of  \eqref{spx} as  the
specified set
% |--------------------------------------------------------------------|
\beq{stsp}
S[u] = \{R(k)=0,\; -\infty<k<\infty;\;p_1=p, \; q_1=q,\,A = 1,\beta=1\},
\end{equation}
% |--------------------------------------------------------------------|
and, correspondingly, the GLM equation becomes
% |--------------------------------------------------------------------|
\beq{glmsp}
W(x,y) + q\,\Gamma\left[\frac{1}{2},\, p\,(x + y)^2\right] +  \zeta\,q\int_{x}^{\infty} \mathrm{d}z \; W(x,z)\,\Gamma\left[\frac{1}{2},\,p\,(z + y)^2\right] = 0, \quad y>x,
\eeq
% |--------------------------------------------------------------------|
with $\zeta\neq0$ a parameter in pursuance of convenience. 
% |--------------------------------------------------------------------|

Since the kernel in \eqref{glmsp} is  not separable, it is a matter of
appositeness to solve \eqref{glm} for $W(x,y)$ in terms of von Neumann
series. 

In that event, consider the initial definition,
\beq{eq22}
K_1(z+y) := \int_z^{\infty} \mathrm{d}z' K(z+z')K(z'+y).
\eeq
hence, after the  GLM equation has been iterated $\mu > 1$ times, there results
% |--------------------------------------------------------------------|
\beq{w_glm}
-W(x,y) =  K(x+y)- \zeta\,\sigma_\mu(x,y) - \rho_\mu(x,y),
\eeq
% |--------------------------------------------------------------------|
where 
% |--------------------------------------------------------------------|
\begin{subequations}
\label{glmsum}
  \begin{align}
    \sigma_\mu(x,y) &= \sum_{\nu=1}^{\mu}\zeta^{\nu-1}\int_{x}^{\infty}\mathrm{d}z\,K_\nu(z+y)K(x+z), \label{glmsum_a} \\
    \rho_\mu(x,y) &= \zeta^{\mu + 1} \int_{x}^{\infty}\mathrm{d}z\,K_{\mu+1}(z+y)W(x,z),\label{glmsum_b}
  \end{align}
\end{subequations}
% |--------------------------------------------------------------------|
and the $\mu$--th  iterated kernel  
\beq{mukernel}
K_\mu(z+y) = \int_{z}^{\infty}\mathrm{d}z'\,K_{\mu-1}(z+z')K(z'+y).
\eeq
% |--------------------------------------------------------------------|

% |--------------------------------------------------------------------|
As the latter prescribes, provided  that the function $K_\mu$ suffices
the Picard iteration, then it  is assured the existence and uniqueness
of the solution to \eqref{glm} by Lindel\"of's apophthegm, thusly, the
convergence of  the von  Neumann series  \eqref{glmsum} remains  to be
tested.
% |--------------------------------------------------------------------|

%	------------------------------------------------------------
%	\\\\\\\\\\\\\\\\\\\\\\\\\\\\\\\\\\\\\\\\\\\\\\\\\\\\\\\\\\\\
%	------------------------------------------------------------

\subsection{Analysis of convergence}
\label{s:convergence}

% |--------------------------------------------------------------------|
The expansion \eqref{w_glm} comprises that  if $\vert K(x+y) \vert$ is
bounded by a number $\epsilon\in\mathbb{R}$  in any closed interval in
$x$ of lenght  $l$, hence, inequity $\vert  K(x+y)\vert \leq \epsilon$
implies that $K_\mu(x+y)$ is also bounded for all $\mu\geq2$, that is,
$\vert K_\mu(x+y)\vert\leq \epsilon (\epsilon\,l)^{\mu-1}$. Therefore,
each term of \eqref{glmsum_a} satisfies
% |--------------------------------------------------------------------|
\begin{equation}
\left\vert \zeta^{\mu-1}\int_{x}^{\infty}\mathrm{d}z\,K_\mu(z+y)K(x+z) \right\vert\leq \epsilon\, (\,\vert\zeta\vert \,\epsilon\,l \,)^{\mu-1} \,\| K(x+z)\|.
\end{equation}
% |--------------------------------------------------------------------|

% |--------------------------------------------------------------------|
In particular,  when  $\vert\zeta\vert <  1$,  the
  sequence $\{\sigma_\mu(x,y)\}$ of partial  sums is a Cauchy sequence
  for  any  positive   real  number  $\delta$  iff
% |--------------------------------------------------------------------|
\begin{equation*}
\begin{split}
\vert \sigma_\nu(x,y) - \sigma_\mu(x,y) \vert & \leq \epsilon \left[\sum_{\kappa=\mu+1}^{\nu} (\, \vert\zeta\vert \,\epsilon\,l \,)^{\kappa-1}\right] \|K(x+y)\| \\
& \leq  \epsilon \,\frac{(\, \vert\zeta\vert \,\epsilon\,l \,)^{\mu} }{1-\vert\zeta\vert \,\epsilon\,l}\, \| K(x+y) \|\\
&\leq \delta,
\end{split}
\end{equation*}
% |--------------------------------------------------------------------|
is held in the very limit $\mu\rightarrow\infty$. Quite in the same way for
% |--------------------------------------------------------------------|
\begin{equation}
\vert\rho_\mu(x,y)\vert \leq \epsilon\, (\,\vert\zeta\vert \,\epsilon\,l\, )^{\mu}\, \|W(x,z)\|,
\end{equation}
% |--------------------------------------------------------------------|
which  means  that $\rho_\mu(x,y)$  will  vanish  uniformly along  the
interval  $l$ while  $\mu\rightarrow\infty$.  Thereupon, the  sequence
$\sigma_\mu(x,y)$  of  continuous  functions  converges  absolute  and
uniformly to the function
% |--------------------------------------------------------------------|
\begin{equation}
\begin{split}
\sigma(x,y) &= \sum_{\mu=1}^{\infty}\zeta^{\mu-1}\int_{x}^{\infty}\mathrm{d}z\,K_\mu(z+y)K(x+z)\\
&= \int_{x}^{\infty}\mathrm{d}z\,\left[\sum_{\mu=1}^{\infty}\zeta^{\mu-1}K_\mu(z+y)\right]K(x+z)\\
&= \int_x^{\infty}\mathrm{d}z\, \Xi(z,y;\zeta) K(x+z),
\end{split}
\end{equation}
% |--------------------------------------------------------------------|
given $\Xi(z,y;\zeta)$  as the  {\em dissolvent kernel},  analogous to
the  higher order  deformation constraint  \eqref{deformation} in  the
ambiance     of    homotopy     analysis     studied    in     section
\ref{s:approximations}.
% |--------------------------------------------------------------------|

% |--------------------------------------------------------------------|
Thus, the solution to \eqref{glm} has the form
% |--------------------------------------------------------------------|
\begin{equation}
-W(x,y) = K(x+y)-\zeta\,\sigma(x,y),\quad y>x,
\end{equation}
% |--------------------------------------------------------------------|
and, on account of that, from \eqref{glmsp} we obtain
% |--------------------------------------------------------------------|
\begin{equation}
\label{sol_glmsp}
- W(x,y) = q\,\Gamma\left[\frac{1}{2},\, p\,(x + y)^2\right] -  q \,\zeta\int_{x}^{\infty}\mathrm{d}z\,\Xi(z,y;\zeta)\,\Gamma\left[\frac{1}{2},\, p\,(x + z)^2\right],
\end{equation}
% |--------------------------------------------------------------------|
whilst the solution  to equations \eqref{spx} is  computed through the
formulae:
% |--------------------------------------------------------------------|
\beq{formulae}
\begin{split}
w(x) &= 2 W(x,x), \\
w(x) &= \int_{x}^{\infty}\mathrm{d}y\, u(y),\\
u(x) &= w_x(x).
\end{split}
\eeq
% |--------------------------------------------------------------------|

% |--------------------------------------------------------------------|
In our particular  case, we claim for a real  valued, regular function
$u(x)$ that vanishes asymptotically, exponentially,
% |--------------------------------------------------------------------|
\beq{gvanish}
\lim_{x\rightarrow\pm\infty}[\,u(x)\exp(\pm2\delta^{(\pm)}x)\,]=0,\quad 0<\delta^{(\pm)}.
\eeq
% |--------------------------------------------------------------------|

% |--------------------------------------------------------------------|
Inasmuch as this scenario holds,  the reflection coefficient $R(k)$ is
meromorphic in the {\em Bargmann strip}
% |--------------------------------------------------------------------|
\begin{equation*}
-\min(\delta^{(-)},\delta^{(+)})<\mathrm{Im}\,k<\delta^{(+)},
\end{equation*}
% |--------------------------------------------------------------------|
where  there is  a  bijection  between the  poles  of  $R(k)$ and  the
discrete                eigenvalues               $k_\alpha=ip_\alpha$
\cite[pp.  68--79]{calogero82}.  For  all  these poles  and  all  the
discrete eigenvalues, the correspondence
% |--------------------------------------------------------------------|
\beq{relax}
\lim_{k\rightarrow \mathrm{i}p_\alpha} [\,(k-\mathrm{i} p_\alpha)R(k)\,] = \mathrm{i} q_\alpha,
\eeq
% |--------------------------------------------------------------------|
is prevailed. 
% |--------------------------------------------------------------------|

% |--------------------------------------------------------------------|
On the other hand, as we  have settled in \eqref{stsp}, the reflection
coefficient $R(k)$ is demanded to be  zero for all $k$, thus, the last
relationship  is  trivial  and,  whether  or  not  our  solutions  are
normalizable, it constitutes a relaxed constraint.
% |--------------------------------------------------------------------|

%	------------------------------------------------------------
%	\\\\\\\\\\\\\\\\\\\\\\\\\\\\\\\\\\\\\\\\\\\\\\\\\\\\\\\\\\\\
%	------------------------------------------------------------

\subsection{Sketching the solutions}
\label{s:dissolvent}

% |--------------------------------------------------------------------|
In general, the dissolvent kernel $\Xi(z,y;\zeta)$ can be built as the
quotient of two functions $P(z,y;\mu)$  and $Q(\mu)$ prescribed by the
series expansion:
% |--------------------------------------------------------------------|
\beq{res_form}
\begin{split}
P(z,y;\zeta) &= \sum_{\mu=0}^{\infty} \frac{(-\zeta)^\mu }{\mu!} \Lambda_\mu(z,y),\\
Q(\zeta) &= \sum_{\mu=0}^{\infty} \frac{(-\zeta)^\mu }{\mu!} \lambda_\mu,\\
\end{split}
\eeq
% |--------------------------------------------------------------------|
with  the  proviso  of  initial data  $\Lambda_0(z,y)=K(z  +  y)$  and
$\lambda_0  = 1$.  As for  the coming  terms $\mu\geq1$,  the function
$\Lambda_\mu(z,y)$ is computed by means of the recursive relation:
% |--------------------------------------------------------------------|
\beq{alg1}
\Lambda_\mu(x,z) = \lambda_\mu\, K(z + y) - \zeta \int_{x}^{\infty} \mathrm{d}s\, K_{\mu-1}(z+s)\,\Lambda_{\mu-1}(s,y),
\eeq
% |--------------------------------------------------------------------|
with
% |--------------------------------------------------------------------|
\beq{alg2}
\lambda_\mu = \int_{l}\mathrm{d}z\, \Lambda_{\mu-1}(z,z),
\eeq
% |--------------------------------------------------------------------|
recalling that $l\in\mathrm{Re}\,\mathscr{X}$.
% |--------------------------------------------------------------------|

% |--------------------------------------------------------------------|
After the  algorithm \eqref{alg1}--\eqref{alg2} has been  iterated one
time, fixing  $\zeta=1$ without  loss of  generality, the  integral in
\eqref{sol_glmsp} is explicitly calculated as
% |--------------------------------------------------------------------|
\begin{equation*}
\begin{split}
& \int_{x}^{\infty}\mathrm{d}z\,\Xi(z,y;1)\,\Gamma\left[\frac{1}{2},\, p\,(x + z)^2\right] = \\
& - q \,(x+y) \,\Gamma^2\left[\frac{1}{2},\, p\,(x + y)^2\right] + \frac{q}{p\sqrt{\pi}} \,\Gamma\left[\frac{1}{2},\, p\,(x + y)^2\right]\,\exp-p^2\,(x+y)^2 \\
& + \frac{2 q p}{\sqrt{\pi}}\int_{x}^{\infty}\mathrm{d}z\,\exp-p^2(x+z)^2\left\{ (z+y)  \, \Gamma\left[\frac{1}{2},\, p\,(z + y)^2\right]  - \frac{q}{p\sqrt{\pi}} \exp-p^2(z+y)^2 \right\},
\end{split}
\end{equation*}
% |--------------------------------------------------------------------|
this    integral    must   converge    to    zero    in   the    limit
$\mu\rightarrow\infty$ of iterations since $\sigma(x,y)$ satisfies the
H\"older  condition  for some  positive  real  constant $\delta$  when
becomes  characterized  as  from  \eqref{glmk}.  Then,  as  it  decays
monotonically  with  each  iteration,   the  same  reasons  stated  in
\ref{s:convergence} are concedable. As a result we drop out the second
term   from  \eqref{sol_glmsp}   and,  therefore,   the  solution   to
\eqref{glmsp} is simply
% |--------------------------------------------------------------------|
\beq{wsp2}
-W(x,y) = q\, \Gamma\left[\frac{1}{2},\, p\,(x + y)^2\right].
\eeq
% |--------------------------------------------------------------------|

% |--------------------------------------------------------------------|
Now,  it is  admitedly suggested  that  by the  direct application  of
formulae \eqref{formulae} to \eqref{wsp2},  the solution to the system
\eqref{spx} has the general form
% |--------------------------------------------------------------------|
\begin{equation}
\label{ansatz}
u(x) = 2\,q\,E(x,x_0)\, \exp[\,-p \,c_1 \,(x \pm x_0)^2\,] + c_2,
\end{equation}
% |--------------------------------------------------------------------|
where  $E(x,x_0)$  is  a  real   valued,  continuous  function  to  be
determined  (as we  will see  in section  \ref{s:detach}), $x_0$  is a
point  in  the neighbourhood  of  $x$  and $c_1,c_2$  are  integration
constants.
% |--------------------------------------------------------------------|

% |--------------------------------------------------------------------|
Furthermore, the function $E(x,x_0)$ must suffice (by itself) equation
\eqref{schro} and,  whenever it vanishes faster  than \eqref{gvanish},
namely
% |--------------------------------------------------------------------|
\begin{equation*}
\lim_{x\rightarrow\infty}[\,E(x,x_0)\exp(\delta x)\,]=0,\quad \delta=\min(\delta^{(-)},\delta^{(+)}),
\end{equation*}
% |--------------------------------------------------------------------|
the  relationship  \eqref{relax}  holds;  consequently  $u(x)$  to  be
normalizable is  not a necessary condition  and the factor $q$  is not
relevant in our quest for exact solutions.
% |--------------------------------------------------------------------|

%	 ____  ____  ____  __    ___  _  _   __   ____  __    ____ 
%	(    \(  __)(_  _)/ _\  / __)/ )( \ / _\ (  _ \(  )  (  __)
%	 ) D ( ) _)   )( /    \( (__ ) __ (/    \ ) _ (/ (_/\ ) _) 
%	(____/(____) (__)\_/\_/ \___)\_)(_/\_/\_/(____/\____/(____)

\section{Detachable solutions}
\label{s:detach}

% |--------------------------------------------------------------------|
Let us  consider the  solution \eqref{ansatz} and  let us  assume that
$E_1=E_1(x,x_0)$  and  $E_2=E_2(x',x'_0)$  satisfy by  themselves  the
nonlinear system \eqref{spx}.  We can look for  separable solutions by
starting  with the  stationary  Schr\"odinger equation  \eqref{schro},
from which the product $E_1 E_2$ has to be solvable, {\em i.e.} 
% |--------------------------------------------------------------------|
\beq{separable}
\left[-\partial^2_x - \partial^2_{x'} +z\, \phi_{12}(\,\vert E_1 E_2 \vert^2\,)\right]E_1 E_2 = \omega\,E_1 E_2,
\eeq
% |--------------------------------------------------------------------|
such that both $\phi=\phi_{12}=\phi_{21}$  and $\phi = \phi_1+\phi_2 =
\phi_2+\phi_1$ must be fulfilled. The  general solutions to the latter
functional equation are $\phi = -a\ln \vert E \vert^2$ and $\phi_\nu =
- a\ln  \vert  E_\nu   \vert^2$,  for  some  real   constant  $a$  and
$\nu=1,2$. More than that, it is required from \eqref{ansatz} that $E$
satisfies simultaneously the Poisson equation, that means
% |--------------------------------------------------------------------|
\beq{poissonx}
(-a\ln\vert E\vert^2)_{xx} = \vert E \vert^2.
\eeq
% |--------------------------------------------------------------------|

% |--------------------------------------------------------------------|
In  spirit of  section \ref{s:approximations},  assume that  $E$ is  a
continuous  real   function  which   behaves  according  to   the  law
\eqref{levi}, hence \eqref{poissonx} has a family of parametric solutions $E\rightarrow E(\cdot;\xi_1,\xi_2)$, namely,
% |--------------------------------------------------------------------|
\beq{spoisson}
E(x,x_0;\xi_1,\xi_2) = 4ap^2 \exp[\,p\,(x \pm x_0)\,] \times \begin{dcases} 
\frac{\exp(p\,\xi_1)}{2 a p^2 + \exp[\,2p(x\pm x_0+\xi_1)\,]}, & \xi_1\in\mathbb{R},  \\
\frac{\exp(p\,\xi_2)}{1 + 2 a p^2\exp[\,2p\,(x\pm x_0+\xi_2)\,]},& \xi_2\in\mathbb{R}, 
\end{dcases}
\eeq
% |--------------------------------------------------------------------|
with $\xi_1,\xi_2$ the real parameters which characterize the solutions.  Notice that, in the very particular case $\xi_1= (2p)^{-1}\ln(2ap^2)$  and $\xi_2 = -(2p)^{-1}\ln(2ap^2)$,
the expression in \eqref{spoisson} is reduced to
% |--------------------------------------------------------------------|
\beq{sech}
E(x,x_0;\xi_1,\xi_2) = \begin{dcases} 
p\sqrt(2a)\, \mathrm{sech}[\,p(x\pm x_0)\,], & \text{for } \xi_1, \\
p\sqrt(2a)\, \mathrm{sech}[\,p(x\pm x_0)\,], & \text{for } \xi_2.
\end{dcases}
\eeq
% |--------------------------------------------------------------------|

% |--------------------------------------------------------------------|
Finally, with the aim  of \eqref{ansatz} and \eqref{spoisson}, and recalling
the   form   of   our  initial   ansatz   $\psi(x,t) \rightarrow \psi(x,t;\xi_1,\xi_2)=u(x;\xi_1,\xi_2)\exp(\pm \mathrm{i}\omega t)$,  we write  down  the  general stationary family of solutions to the nonlinear system \eqref{spx} as
% |--------------------------------------------------------------------|

\beq{solution}
\psi(x,t;\xi_1,\xi_2) = 2\,q\,E(x,x_0;\xi_1,\xi_2)\, \exp[\,-p \,c_1 \,(x \pm x_0)^2\,] \exp(\pm \,\mathrm{i}\omega t) + c_2,
\eeq

% |--------------------------------------------------------------------|
which are Schr\"odinger--Poisson,  stationary, single, parametric solitons subject
to the conservative dispersion relation:
% |--------------------------------------------------------------------|
\beq{dispersion}
\omega = \frac{-p^2 + 12 a p^4 - 
 4 a^2 p^6 + 2\,(1 + 2 a p^2)^2 \left[ p - 
    a b \ln\left(\frac{8 a p^2}{1 + 2 a p^2}\right)\right]}{(1 + 2 a p^2)^2},
\eeq
% |--------------------------------------------------------------------|
where we have neglected $\xi_1,\xi_2$ and $x_0$ for a tight writing. 
% |--------------------------------------------------------------------|

% |--------------------------------------------------------------------|
The wave function in \eqref{solution} corresponds to a {\em quiescent}
soliton centred at $x_0$, oscillating with speed $\omega$ and decaying
according to the eigenvalue $p$ associated to a single bounded state.
% |--------------------------------------------------------------------|

% |--------------------------------------------------------------------|
In  case $A\neq1$  in  \eqref{stsp}, we  would  obtain the  equivalent
multiple--stationary  soliton  solutions  once applied  the  procedure
described above.  This is not an  easy task: the more  bounded states,
the more demanding to solve a system  of $A$ GLM equations in a closed
form. For this  purpose, a numerical approach might  be an alternative
to achieve the results.
% |--------------------------------------------------------------------|

% |--------------------------------------------------------------------|
As a final remark, note that the number of discrete eigenvalues $A$ of
the  Schr\"odinger operator  ${\bf  H}$ in  \eqref{schro} is  computed
through a function $J=J(x)$ defined by the nonlinear equation
% |--------------------------------------------------------------------|
\beq{number}
c\,u\cos^2J - c^{-1}\,\sin^2J + J_x = 0,
\eeq
% |--------------------------------------------------------------------|
subject to the boundary condition 
% |--------------------------------------------------------------------|
\beq{numboundary}
J(-\infty)=0,
\eeq
% |--------------------------------------------------------------------|
$c$ standing for any positive constant, then 
% |--------------------------------------------------------------------|
\beq{numformula}
A=\lfloor J(\infty)/\pi\rfloor. 
\eeq
% |--------------------------------------------------------------------|

% |--------------------------------------------------------------------|
Certainly, $\omega$ is discretized since $a$ and $b$ are constants and
$p$ is  a discrete eigenvalue of  ${\bf H}$, that is,  because we have
considered  a single  soliton  ($A=1$)  there is  a  single value  for
$\omega$. In  consequence, for  these particular  case, \eqref{ansatz}
must     satisfy      \eqref{number}     such      that     $1=\lfloor
J(\infty)/\pi\rfloor$, just as it  is. Properly, each multiple soliton
solution to \eqref{sp} has to  fulfill \eqref{number} so the number of
nonlinear  superpositions  (or  Bianchi permutations)  coincides  with
\eqref{numformula}.
% |--------------------------------------------------------------------|

\section{Discussion}
\label{s:discussion}

% |--------------------------------------------------------------------|

We  presented  exact,  stationary, parametric solutions  to  the
Schr\"odinger--Poisson  nonlinear   system  of   partial  differential
equations. In the first part of our study, we showed how the homotophy analysis method suffices to integrate the system \eqref{sp}. However, we found that to compute the integrals in \eqref{sol} is not straightforward, but instead, a limitation carried with this programme. As obvious, this approach might provide first numerical computations.   

In the second part of our study, we found a family of exact solutions to the Schr\"odinger--Poisson system through the inverse scattering method. The inherent relation between these two techniques has been tangentially depicted. 

In both cases, we addressed our discussion to the theory of integral equations, either nonlinear and linear. To solve these equations is not an easy task, nonetheless, the techniques come up with a deep understanding about the very nature of the problem in itself. At this point, we have not been able to seek for exact solutions by any algebraic method. Also, without counting the inverse scattering method, the mainstream techniques described in soliton theory are insufficient to solve our problem in a closed form. 

Our solutions just cover the stationary configuration of the Schr\"odinger--Poisson system and are related to a single eigenvalue of the Schr\"odinger operator. To seek for closed solutions out of equilibrium is not a goal we have achieved yet. Quite the same for multiple Bianchi permutations, {\em i.e.}  more than one eigenvalue (bounded state) of the Schr\"odinger operator. In that case, we should find something at the same level of a B\"acklund (or Darboux) transformation in order to induce a nonlinear superposition of bounded states. 

% |--------------------------------------------------------------------|

\section*{Acknowledgments}

The authors are glad to thank the valuable remarks and discussions by their colleagues at the Department of Physics at CINVESTAV, Mexico City.  

% |--------------------------------------------------------------------|

%	 ____  __  ____ 
%	(  _ \(  )(  _ \
%	 ) _ ( )(  ) _ (
%	(____/(__)(____/

\bibliographystyle{unsrt}
\bibliography{references}
%	-------------------- 
\end{document}